\documentclass[aps,prb,preprint,showkeys,superscriptaddress]{revtex4}

\usepackage{graphicx}		
\usepackage{dcolumn}		
\usepackage{bm}				
\usepackage{amsmath}
\usepackage{amssymb}
\usepackage{amsfonts}
\usepackage{amssymb}
\usepackage{color}
\usepackage{enumitem}
\usepackage{hhline}

\begin{document}

\title{Power transfer in magnetoelectric resonators}

\author{Frederic Vanderveken}
\email[E-mail: ]{frederic.vanderveken@imec.be}
\affiliation{Imec, 3001 Leuven, Belgium}
\affiliation{KU Leuven, Departement Materiaalkunde, SIEM, 3001 Leuven, Belgium}
\author{Bart Sor\'ee}
\affiliation{Imec, 3001 Leuven, Belgium}
\affiliation{KU Leuven, Departement Elektrotechniek, TELEMIC, 3001 Leuven, Belgium}
\author{Florin Ciubotaru}
\affiliation{Imec, 3001 Leuven, Belgium}
\author{Christoph Adelmann}
\email[E-mail: ]{christoph.adelmann@imec.be}
\affiliation{Imec, 3001 Leuven, Belgium}

\begin{abstract}
We derive an analytical model for the power transfer in a magnetoelectric film bulk acoustic resonator consisting of a piezoelectric--magnetostrictive bilayer. The model describes the dynamic magnetostrictive influence on the elastodynamics via an effective frequency-dependent stiffness constant. This allows for the calculation of both the magnetic and elastic power absorption in the resonator as well as of its energy efficiency when such a resonator is considered as a magnetic transducer. The model is then applied to example systems consisting of piezoelectric ScAlN and magnetostrictive CoFeB, Ni, or Terfenol-D layers. 
\end{abstract}


\maketitle

\section{Introduction}

In recent years, there has been an increasing interest in magnetoelectric transducers that efficiently convert electrical signals to magnetic excitations and \emph{vice versa}. Especially in the emerging field of magnonics,\cite{chumak_magnon_2015,  mahmoud_introduction_2020, barman_2021_2021,chumak_advances_2022} which uses spin waves as information carriers, magnetoelectric transducers are thought to be critical elements to enable ultralow-power data processing.\cite{khitun_non-volatile_2011,dieny_opportunities_2020, mahmoud_introduction_2020} To date, several types of spin-wave transducers have been proposed and realized, such as inductive antennas\cite{Demidov2009, Chumak2009, Ciubotaru2016, connelly_efficient_2021, vanderveken2022} as well as transducers based on spin-orbit\cite{Divinskiy2018,Talmelli2018} or spin-transfer torques.\cite{Madami2011} However, all these mentioned transducers are current-based and therefore inevitably have associated Joule heating, which strongly reduces their energy efficiency when miniaturized.\cite{vanderveken2022}

Another type of magnetoelectric transducer utilizes the combination of the piezoelectric and magnetostrictive effects. Such transducers employ voltage signals instead of currents, and therefore are expected to operate at a much higher energy efficiency at nanoscale dimensions.\cite{khitun_non-volatile_2011, mahmoud_introduction_2020, khitun_magnetoelectric_2009} In these transducers, elastodynamics are generated from AC voltage signals via the piezoelectric effect, which then couple to the magnetization via the magnetoelastic effect. Therefore, much research has recently been devoted to the question how the magnetization dynamics dynamically couple to the elastodynamics.\cite{Bhuktare17,Verba2021,Babu2021,Polzikova2016,Alekseev2020, Bichurin2010, Litvinenko2021}

Several approaches to generate magnetization dynamics via dynamic strain have been pursued. Most recent work has focused on the coupling of magnetization dynamics to surface acoustic waves (SAWs) that are generated in piezoelectric substrates by interdigitated transducers.\cite{,Cherepov2014,Bhuktare17, Foerster2017, Kittmann2018, Castilla20, Verba2021} While the magnetoelastic coupling can in principle be efficient in such structures,\cite{Mazzamurro2020} reaching the multi-GHz frequencies required for strong coupling is challenging. In addition, SAW devices are difficult to scale to nm dimensions, which are required for scaled spintronic applications.\cite{dieny_opportunities_2020, mahmoud_introduction_2020} By contrast, devices based on bulk acoustic waves have received less attention so far despite perspectives for higher frequency operations as well as better scalability. Such devices consist of a piezoelectric--magnetostrictive bilayer that can locally generate elasto- and magnetodynamics by application of a voltage.\cite{Dutta2014,Pertsev2008, Balinskiy2018,Vanderveken2020} 

In all cases, the magnetoelectric coupling can benefit from mechanical resonances of the system. As an example, high overtone bulk acoustic resonators (HBARs) incorporating an additional magnetostrictive layer inside the device have been modeled and studied experimentally.\cite{Polzikova2016,Polzikova2018, Polzikova2018b, Polzikova2019, Polzikova2019b,Polzikova2019c, Alekseev2020, Bichurin2010, Litvinenko2021} A key result was that the magnetoelectric coupling was strongly enhanced at HBAR resonances. However, from a practical perspective as spintronic transducers, film bulk acoustic resonators (FRBARs) appear most suitable, for example as solidly mounted resonators (SMRs).\cite{lakin_solidly_1995,weinstein_resonant_2010} Indeed, such devices have been proposed as \emph{e.g.}~spin-wave transducers for magnonic logic applications.\cite{Khitun_2007,Khitun11}  While the basic physical behavior has been modeled before,\cite{Bichurin2010,Bichurin2012} these studies have focused on the magnetoelastic coupling and the resonance effect. However, they did not include a description of the power flow, device scalability, and transducer efficiencies. Hence, although magnetoelectric FBAR-like transducers have been proposed as efficient building blocks of spintronic and magnonic logic circuits,\cite{Khitun_2007,Khitun11} their efficiency and its the dependence on material parameters, as well as their scaling characteristics are not yet understood.

In this work, we present a theory that describes the magnetoelastic coupling and the power flow in a piezoelectric--magnetostrictive thin bilayer FBAR at concurring mechanical and magnetic resonances. Analytical expressions for both the elastic and magnetic power losses as well as the power transfer into the magnetic domain allow for a discussion of the transducer efficiency. Finally, a case study illustrates the influence of the material and dimensional parameters on the transducer performance. 

\section{Resonator geometry and configuration}

\begin{figure}[tbh]
	\includegraphics[width=8.5cm]{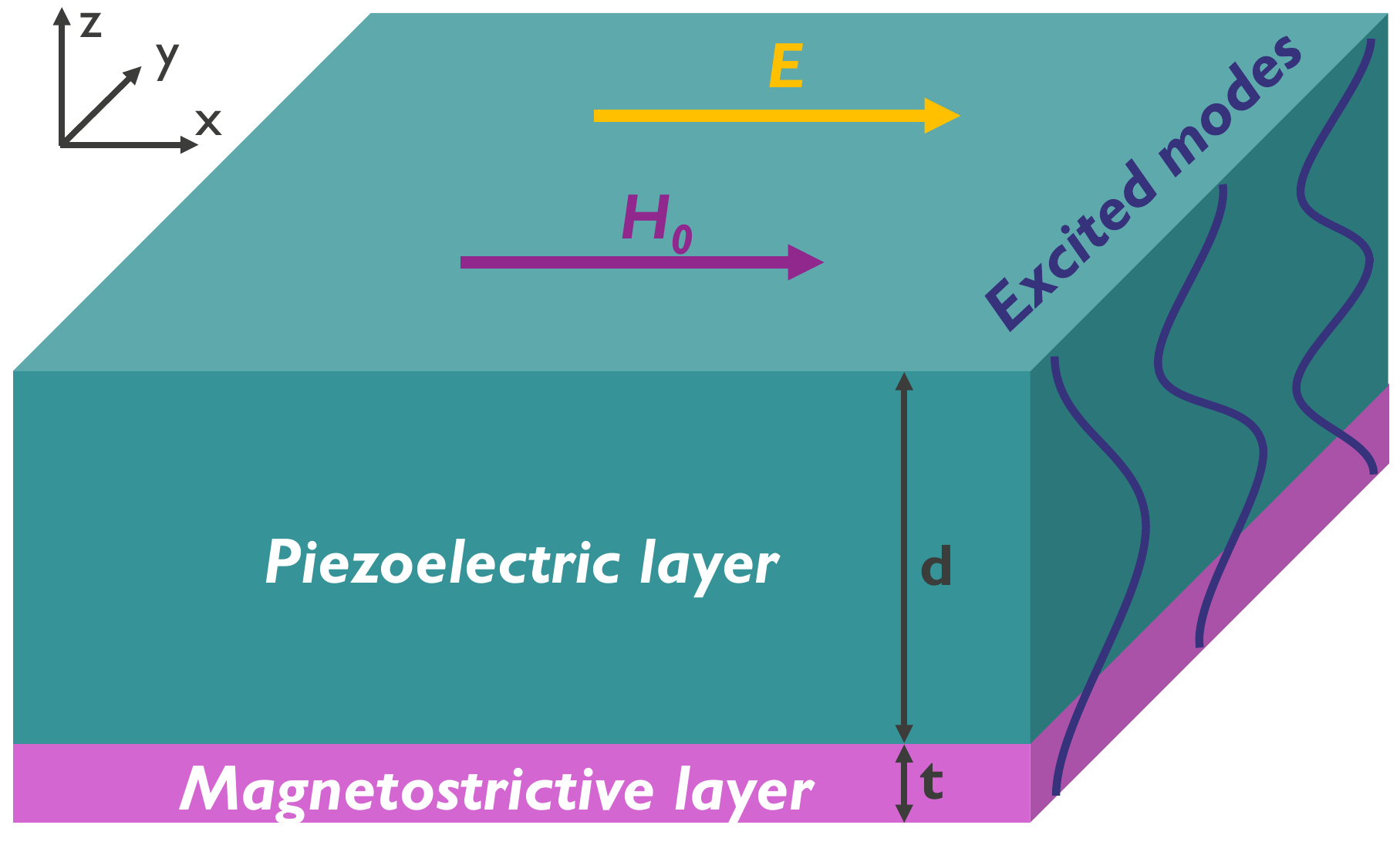}
	\caption{Structure of the magnetoelectric resonator and the coordinate system. $E$ is the time-varying applied electric field and $H_0$ is the static external magnetic field.}
	\label{fig:structure}
\end{figure}

The studied structure consists of a piezoelectric--magnetostrictive bilayer with thickness $d + t$, as depicted in Fig. \ref{fig:structure}. The lateral dimensions are assumed to be much larger than the thickness, which allows the structure to be approximated as a 1D system with spatial variation of magnetization and displacement only in the $z$-direction. We further assume free elastic boundary conditions at the top and bottom of the resonator, similar to an ideal FBAR device.\citep{Rosenbaum1988} The resonator can be excited by applying a time-dependent electric field across the piezoelectric element, resulting in dynamic strain generation inside the bilayer. The strain extends into the magnetostrictive layer and consequently also excites magnetization dynamics. 

There are several different configurations for the direction of magnetization and applied electric field. Some configurations give only weak second order coupling between the elastic and magnetic domain, whereas others have strong first order coupling. In this paper, we study the case where the electric field is applied laterally along the $x$-direction parallel to the static magnetization, as illustrated in Fig. \ref{fig:structure}. This configuration allows for confined shear elastic wave resonator modes along the structure thickness, leading to efficient first-order magnetoelastic coupling. \cite{Duflou2017, Vanderveken2020}

\section{Boundary value problem}

In this section, the differential equations describing the dynamics inside the piezoelectric and magnetostrictive medium are derived. Subsequently, the boundary conditions are specified and an overall solution of the boundary value problem is presented. This solution will be used to calculate the magnetic and elastic power absorption as well as the transducer efficiency in the following sections.

\subsection{Equations of motion in the piezoelectric medium}

The constitutive relations in the piezoelectric medium are given by\cite{Rosenbaum1988, IEEE}
\begin{align}
\label{eq:constitutive_p}
    \sigma_{ij} &= \sum_{k,l}c^E_{ijkl} S_{kl} - \sum_{k}e_{kij} E_k\\
    D_i &= \sum_{k,l} e_{ikl}S_{kl} + \sum_{j} \varepsilon^S_{ij}E_j \,.
\end{align}
Here, $\sigma_{ij}$ is the stress tensor and $c_{ijkl}^E$ the stiffness tensor under constant electric field. $S_{ij}$ is the strain tensor and $e_{kij}$ the piezoelectric coefficient tensor (with unit C/m$^2$). $E_i$ is the electric field, $D_i$ the electric displacement, and $\varepsilon^S_{ij}$ the permittivity tensor under constant strain. 

For a lateral electric field in the $x$-direction and a piezoelectric medium with tetragonal or hexagonal crystal symmetry (\emph{e.g.}, ScAlN used in the case study), the piezoelectrically induced displacement is parallel to the electric field direction and thus also along the $x$-direction.\citep{Rosenbaum1988} Moreover, since the system is spatially uniform in the lateral directions $x$ and $y$, the $xz$ shear strain is the only nonzero component as the only nonzero displacement gradient is in the $z$-direction, \emph{i.e.} $S_{xz} = S_{zx} = \frac{1}{2}\frac{\partial u_x}{\partial z}$. In this case, according to Eq. \eqref{eq:constitutive_p}, there is also only one nonzero stress component given by
\begin{equation}
\label{eq:stress_p}
\sigma_{xz}= \sigma_{zx} = c_{55}^E \times 2S_{xz} - e_{15} E_x , 
\end{equation}
with $c_{55}^E$ the shear stiffness constant under constant electric field, $e_{15}$ the piezoelectric coupling constant, and $E_x$ the applied electric field. In the following, the indices will be omitted and replaced by the superscript $^p$ for variables in the piezoelectric and $^m$ for variables in the magnetostrictive medium.

The elastodynamic equation of motion for the piezoelectric layer is then given by
\begin{equation}
\label{eq:elastodynamic_p}
    \rho^p \frac{\partial^2 u^p}{\partial t^2} = f^p_\mathrm{el} ,
\end{equation}
with $\rho^p$ the mass density of the piezoelectric medium, $u^p \equiv u_x$ the displacement, and $f^p_\mathrm{el}$ the elastic body force given by
\begin{equation}
\label{eq:fel_p}
    f^p_\mathrm{el} = \frac{\partial }{\partial z}\sigma^p =  \frac{\partial }{\partial z} \left(2c^pS^p -eE\right) = c^p \frac{\partial^2 u^p}{\partial z^2} \,.
\end{equation} 
Substituting this expression into the equation of motion in Eq.~\eqref{eq:elastodynamic_p} results in the wave equation
\begin{equation}
\label{eq:wave_p}
    \frac{\partial^2 u^p}{\partial t^2} = v^p \frac{\partial^2 u^p}{\partial z^2} ,
\end{equation}
with phase velocity $v^p = \sqrt{\frac{c^p}{\rho^p}}$ and dispersion relation $ \omega = v^pk$. Mechanical losses in the piezoelectric medium can be incorporated by a complex stiffness constant $c^p = c^p_r+ic^p_i$. The real part of the stiffness constant determines the wavelength of the acoustic wave and its dispersion relation, whereas the complex part describes the its damping. For acoustic excitations, especially in FBARs, the imaginary part is typically several orders of magnitude lower than the real part.\cite{Rosenbaum1988}

The most common loss mechanism at GHz frequencies in piezoelectric media is viscoelastic damping, for which the imaginary part of the stiffness can be written as\cite{Rosenbaum1988, Soh2012} 
\begin{equation}
    c^p_i = \omega \zeta
\end{equation}
with $\omega$ the angular frequency and $\zeta$ the viscosity coefficient of the medium. However, to describe acoustic resonators, an effective loss factor is often employed, \emph{i.e.}~the $Q$-factor, which is more easily accessible experimentally. The $Q$-factor can be related to the complex stiffness constant via\cite{Rosenbaum1988}
\begin{equation}
\label{eq:ci_p}
    Q = \frac{c^p_r}{c^p_i} \,.
\end{equation}

When losses are considered, the wavenumber becomes a complex quantity and is then given by 
\begin{equation}
    k^p = \omega\sqrt{ \frac{ \rho^p}{c^p_t}} =\omega\sqrt{ \frac{ \rho^p}{c^p_r + ic^p_i}}\,.
\end{equation}
As a result, also the phase velocity becomes complex and can be written as $v^p = v^p_r + iv^p_i$. The root of a complex number is given by
\begin{equation}
\sqrt{z} = \sqrt{r}\frac{z+r}{|z+r|}
\end{equation}
with $r$ the modulus of the complex number. Then, with $c^p_i\ll c^p_r$, the effective complex wavenumber can be rewritten as
\begin{equation}
\label{eq:kt_p}
k^p \approx   k^p_r -i k^p_i 
\end{equation}
with
\begin{equation}
\label{eq:ki_p}
    k^p_i = \frac{c^p_i}{2c^p_r} k^p_r = \frac{k^p_r}{2Q}\,.
\end{equation}

\subsection{Equations of motion in the magnetostrictive film}

In a magnetostrictive layer, there is an additional magnetoelastic body force $f_\mathrm{mel}$ in addition to the elastic body force $f^m_{el}$. Hence, the elastodynamic equation of motion inside the magnetostrictive layer becomes
\begin{equation}
\label{eq:elastodynamic_m}
    \rho^m \frac{\partial^2 u^m}{\partial t^2} = f_\mathrm{el}^m +  f_\mathrm{mel} \,.
\end{equation}
As above, the elastic body force is given by
\begin{equation}
\label{eq:fel_m}
f_\mathrm{el}^m = \frac{\partial }{\partial z}\sigma^m =  \frac{\partial }{\partial z} \left(c^m_0 \times 2S^m\right)= c^m_0 \frac{\partial^2 u^m}{\partial z^2} \, ,
\end{equation}  
\noindent with $c^m_0$ the stiffness constant of the magnetostrictive material in absence of magnetoelastic coupling. The magnetoelastic body force is given by (see appendix A)
\begin{align}
\bm{f}_\mathrm{mel}  &=  2 \frac{B_1}{M_S^2} \begin{bmatrix}
M_x \frac{\partial M_x}{\partial x} \\ M_y \frac{\partial M_y}{\partial y} \\ M_z \frac{\partial M_z}{\partial z} \end{bmatrix} \\
&+ \frac{B_2}{M_S^2} \begin{bmatrix}
M_x\left( \frac{\partial M_y}{\partial y} + \frac{\partial M_z}{\partial z} \right)  + M_y \frac{\partial M_x}{\partial y} + M_z \frac{\partial M_x}{\partial z} \\
M_y\left( \frac{\partial M_x}{\partial x} + \frac{\partial M_z}{\partial z} \right) + M_x \frac{\partial M_y}{\partial x} + M_z \frac{\partial M_y}{\partial z}  \\
M_z\left( \frac{\partial M_x}{\partial x} + \frac{\partial M_y}{\partial y} \right)  + M_x \frac{\partial M_z}{\partial x} + M_y \frac{\partial M_z}{\partial y} 
\end{bmatrix}\,.
\end{align}
Here, $B_1$ and $B_2$ are the magnetoelastic coupling constants of the medium and $M_S$ is the saturation magnetization. Considering only nonzero magnetization gradients along the film thickness in the $z$-direction, \emph{i.e.} only $\frac{\partial m}{\partial z}  \neq0$, leads to
\begin{equation}
    \bm{f}_\mathrm{mel}  = 2 \frac{B_1}{M_S^2} \begin{bmatrix}
0 \\ 0 \\ M_z \frac{\partial M_z}{\partial z} \end{bmatrix} + \frac{B_2}{M_S^2} \begin{bmatrix}
M_x \frac{\partial M_z}{\partial z} + M_z \frac{\partial M_x}{\partial z}  \\
M_y \frac{\partial M_z}{\partial z} +  M_z \frac{\partial M_y}{\partial z}  \\
0
\end{bmatrix}\,.
\end{equation}
Furthermore, in the linear magnetic regime, \emph{i.e.}~for $M_y,M_z \ll M_S$ and $M_x\approx M_S$, second order terms in the magnetization can be neglected. This results in
\begin{equation}
    \bm{f}_\mathrm{mel} \approx  \frac{B}{M_S}\frac{\partial M_z}{\partial z} \bm{\hat{e}}_{x}\,,
\end{equation}
\noindent with $B\equiv B_2$.

Assuming that the magnetostrictive layer is much thinner than the acoustic wavelength, the linearized Landau–Lifshitz–Gilbert (LLG) equation can be used to express the dynamic magnetization in the magnetostrictive layer at angular frequency $\omega$ as a function of the shear strain (see appendix B)
\begin{equation}
\label{eq:M_ifo_S}
    M_z = \frac{2B\gamma  \omega_y}{\omega^2 -\omega_y\omega_z} S^m = \frac{B\gamma  \omega_y}{\omega^2 -\omega_y\omega_z} \frac{\partial u^m}{\partial z}\,.
\end{equation}
Here, $\omega_y = \omega_0+i\omega\alpha$ and $\omega_z = \omega_0+\omega_M+i\omega\alpha$, with $\omega_0 = \mu\gamma H_0$ and $\omega_M = \mu\gamma M_S$. $\gamma$ is the gyromagnetic ratio, $\alpha$ is the Gilbert damping parameter, $\mu$ the permeability, and $M_S$ the saturation magnetization of the magnetostrictive material. $H_0$ is the external applied magnetic field. Substituting this into the magnetoelastic body force leads to
\begin{equation}
\label{eq:fmel_m}
    f_\mathrm{mel} =  \frac{B^2\gamma  \omega_y}{M_S\left(\omega^2 -\omega_y\omega_z\right)}\frac{\partial^2 u}{\partial z^2}\,.
\end{equation}

Substituting the body forces, \emph{i.e.}~Eqs. \eqref{eq:fel_m} and \eqref{eq:fmel_m}, into the equation of motion \eqref{eq:elastodynamic_m} leads again to a wave equation
\begin{equation}
\label{eq:wave_m}
    \frac{\partial^2 u^m}{\partial t^2} = v_t^m \frac{\partial^2 u^m}{\partial z^2} 
\end{equation}
with the phase velocity $  v^m = \sqrt{\frac{c_t^m}{\rho^m}} $ and an effective stiffness of
\begin{equation}
\label{eq:ct_m}
    c^m = c^m_0 +\frac{B^2\gamma  \omega_y}{M_S\left(\omega^2 -\omega_y\omega_z\right)} \equiv c^m_0 + c^m_B\,.
\end{equation}

While the intrinsic mechanical stiffness constant $c^m_0$ is frequency independent, the magnetic contribution $c^m_B$ is complex and frequency dependent. Moreover, Eq.~\eqref{eq:ct_m} indicates that $c^m_B$ has the same frequency response as the dynamic magnetic susceptibility. Thus, the magnetic contribution to the effective stiffness allows for the treatment of the magnetostrictive system as a regular lossy elastic medium with a complex and frequency dependent stiffness.

To understand the additional magnetic loss mechanism in more detail, the complex effective stiffness can be explicitly divided into real and imaginary parts as $c^m \equiv c^m_r+ic^m_i$. Assuming weak magnetic Gilbert damping, \emph{i.e.} $\alpha\ll1$, the real and imaginary part of the effective stiffness become respectively (see appendix B)
\begin{equation}
\label{eq:cr_m}
c^m_r = c^m_0 +\frac{B^2\gamma}{M_S} \frac{ \omega_0(\omega^2 - \omega_\mathrm{res}^2) - \omega^2\alpha^2(2\omega_0+\omega_M)} {(\omega^2 - \omega_\mathrm{res}^2)^2 + \omega^2\alpha^2(2\omega_0+\omega_M)^2 }
\end{equation}
and 
\begin{equation}
\label{eq:ci_m}
c^m_i = \frac{B^2\gamma}{M_S} \frac{ \omega\alpha(\omega_0^2+\omega^2)} {(\omega^2 - \omega_\mathrm{res}^2)^2 + \omega^2\alpha^2(2\omega_0+\omega_M)^2 } \,.
\end{equation}

\noindent with $\omega_\mathrm{res}^m = \Re\sqrt{\omega_y\omega_z} = \sqrt{\omega_0(\omega_0+\omega_M)} $ the magnetic resonance frequency.

For ``conventional'' ferromagnetic materials (\emph{e.g.}~transition metals or ferrites), the magnetic contribution $c^m_B$ to the stiffness is typically much smaller than $c^m_0$. Therefore, the real part of the magnetic contribution can be neglected, and $c^m_r \approx c^m_0$. Similarly, the modulus of the effective stiffness can be approximated by $|c^m_t| = \sqrt{(c^m_r)^2 + (c^m_i)^2} \approx c^m_0 $. Furthermore, similar to the imaginary part of the dynamic magnetic susceptibility, also the imaginary part of the stiffness $c^m_i$ reaches its maximum at the magnetic resonance frequency $\omega_r^m$. This maximum is given by
\begin{equation}
\label{eq:cires_m}
c^m_i(\omega_\mathrm{res}^m) = \frac{B^2\gamma\omega_0}{M_S\omega_\mathrm{res} \alpha(2\omega_0+\omega_M)}
\end{equation}

Analogous to a lossy piezoelectric medium, the complex stiffness of the magnetostrictive medium also results in complex wavenumbers given by
\begin{equation}
\label{eq:kt_m}
    k^m = k^m_r  - i k^m_i 
\end{equation}
with
\begin{equation}
\label{kmr}
k^m_r = \omega \sqrt{\frac{\rho^m}{c^m}}
\end{equation}
and
\begin{equation}
\label{ki_m}
k^m_i = \frac{c^m_i}{2c^m}k^m_r  \,.
\end{equation}

\subsection{Solution of the boundary value problem}

As shown above, the dynamic response of the displacement is described by wave equations in both the piezoelectric and magnetostrictive layers. Hence, a wave ansatz can be used in both layers as a solution to the equations of motion \eqref{eq:wave_p} and \eqref{eq:wave_m}. However, both layers have a different phase velocity and thus also a different wavenumber at a given frequency. Inside the piezoelectric medium, the displacement and strain ansatzes can be written as
\begin{align}
    u^p(z) &= A_1 e^{ik^pz} + A_2 e^{-ik^pz} \\
    \label{eq:Sp_gen}
    S^p(z) &= \frac{ik^p}{2} \left( A_1 e^{ik^pz} - A_2 e^{-ik^pz} \right)\,.
\end{align}

\noindent Following Eq.~\eqref{eq:stress_p}, the total stress in the piezoelectric layer then becomes 
\begin{align}
\sigma^p(z) &= 2c^pS^p(z)-eE \\
&=  ic^p_tk^p \left( A_1 e^{ik^pz} - A_2 e^{-ik^pz} \right) -eE \,.
\end{align}

Inside the magnetostrictive layer, the displacement and strain ansatzes are given by
\begin{align}
    u^m(z) &=  A_3 e^{ik^mz} + A_4 e^{-ik^mz}  \\
    \label{eq:Sm_gen}
    S^m(z) &= \frac{ik^m}{2}  \left(   A_3e^{ik^mz} - A_4e^{-ik^mz} \right)\,.
\end{align}

\noindent Analogous to the total body force, the total stress in the magnetostrictive layer has both elastic and magnetoelastic contributions. The magnetoelastic stress tensor is derived in Appendix A. Considering both contributions, the total stress becomes
\begin{align}
    \sigma^m(z) &= c^m_0(2S^m(z)) +\frac{B}{M_S} M_z \\
    &= 2c^m_0S^m(z) +\frac{2B^2\gamma  \omega_y}{M_S(\omega^2 -\omega_y\omega_z)} S^m(z) \\
    &= ic^mk^m\left( A_3e^{ik^mz} - A_4e^{-ik^mz} \right) \,.
\end{align}

The boundary conditions impose zero stress at the top surface, continuity of displacement and stress at the interface between the piezoelectric and magnetostrictive layers, and zero stress at the bottom surface, \emph{i.e.}
\begin{align}
    &\sigma^p(z=d)=0 \,,\\
    &\sigma^p(z=0)=\sigma^m(z=0) \,,\\
    &u^p(z=0)=u^m(z=0) \,,\\
    &\sigma^m(z=-t)=0 \,.
\end{align}
\noindent These four conditions can be used to find expressions for the four constants $A_1$ to $A_4$. Inserting these expressions into Eqs.~\eqref{eq:Sp_gen} and \eqref{eq:Sm_gen} leads to expressions for the strain inside the device. In the piezoelectric layer, the strain is given by
\begin{widetext}
\begin{equation}
\label{eq:S_original_p}
    S^p(z) =  eE\frac{c^m k^m \sin (k^m t) \cos (k^p z)+2 c^p k^p \sin \left(\frac{d
   k^p}{2}\right) \cos (k^m t) \cos \left(\frac{1}{2} k^p (d-2 z)\right)}{2 c^m c^p
   k^m \cos (d k^p) \sin (k^m t)+2 \left(c^p\right)^2 k^p \sin (d k^p) \cos (k^m
   t)} \equiv eE S^p_k \,,
\end{equation}
whereas in the magnetostrictive layer, the strain is given by
\begin{equation}
\label{eq:S_original_m}
    S^m(z) = eE \frac{ k^m_t \sin^2\left(\frac{d k^p_t}{2}\right) \sin (k^m_t (t+z))}{c^m_t k^m_t \cos(d k^p_t) \sin (k^m_t t)+c^p_t k^p_t \sin (d k^p_t) \cos (k^m_t t)} \,.
\end{equation}
Equation~\eqref{eq:S_original_m} can be further simplified when the magnetic layer is much thinner than the acoustic wavelength, \emph{i.e.}~when $k^m(t+z)\ll 1$. This leads to
\begin{equation}
\label{eq:Sm}
S^m(z)\approx eE (t+z)\frac{ \left(k^m_t\right)^2 \sin^2\left(\frac{d k^p_t}{2}\right) }{D(\omega)} \equiv eE (t+z) S^m_k 
\end{equation}
\noindent with the strain denominator $D(\omega)$ given by
\begin{equation}
D(\omega) = c^m_t k^m_t \cos(d k^p_t) \sin (k^m_t t)+c^p_t k^p_t \sin (d k^p_t) \cos (k^m_t t)\,.
\label{Eq:S_denom}
\end{equation}
Note that the strain in both layers is a complex number since both the stiffness and wavenumber are complex quantities in this model. 
\end{widetext}

\section{Magnetic power transfer}

The strain in the magnetostrictive layer results in a dynamic magnetoelastic field, which in turn induces magnetization dynamics. Hence, a part of the total input energy is transferred into the magnetic domain. The power transfer density into the magnetic system, originating from the magnetoelastic field, is given by \cite{Kobayashi73, Weiler11}
\begin{equation}
    p(z) = \frac{ i\omega}{2} \mu_0  \mathbf{M}_\mathrm{dyn} \cdot \mathbf{H}_\mathrm{mel}^*\,.
\end{equation}
Considering the magnetization as in Eq.~\eqref{eq:M_ifo_S} and the resulting magnetoelastic field (see Eq.~\eqref{eq:Hmel}), the power transfer density becomes
\begin{align}
    p(z) &= i\frac{\omega \mu_0}{2} M_z H_\mathrm{mel,z}^* \\
    &=  i\frac{\omega \mu_0}{2} \left(  \frac{2B\gamma  \omega_y}{\omega^2 -\omega_y\omega_z} S^m \right) \left( \frac{-2B(S^m)^*}{\mu_0M_S} \right) \\
    &= -i\frac{2\omega B^2\gamma\omega_y}{M_S\left( \omega^2 -\omega_y\omega_z\right)}|S^m(z)|^2 \\
    &= -i2\omega c^m_B|S^m(z)|^2 \,.
\end{align}
The power density can again be separated into real and imaginary parts, \emph{i.e.} $p(z) \equiv  p_r(z) + i p_i(z)$, in an analogous way as the effective magnetostrictive stiffness. The power absorption corresponds to the real part of the power density whereas the reactive power oscillation corresponds to the imaginary part of the power density. The total power absorption is obtained by integrating the real part of the power density over the magnetic volume. The power per unit area then becomes
\begin{equation}
    P_\mathrm{m} = \int\limits_{-t}^0 p_r(z) dz = 2\omega c^m_i \int\limits_{-t}^{0}|S^m(z)|^2 dz \,.
\end{equation}

Using Eq.~\eqref{eq:Sm} and performing the integration leads to the total absorbed magnetic power, which is given by
\begin{equation}
\label{eq:Pm}
P_\mathrm{m}(\omega) = 2\omega c^m_i e^2 E^2 \frac{t^3}{3} |S^m_k(\omega)|^2  \,.
\end{equation}
Note that this expression has a strong frequency dependence because of the resonant behavior of both the magnetic and elastic systems. The frequency behavior of the magnetic system is captured in the imaginary part of the stiffness constant, $c^m_i(\omega)$, whereas the frequency behavior of the elastic system is contained in the strain magnitude factor $S^m_k(\omega)$, defined in Eq.~\eqref{eq:Sm}. At ferromagnetic resonance, the power absorption per unit area is given by
\begin{equation}
    P_\mathrm{m}(\omega_\mathrm{res}^m) =\frac{2}{3} E^2 t^3 \frac{e^2B^2\gamma\omega_0}{M_S \alpha(2\omega_0+\omega_M)}|S^m_k(\omega_\mathrm{res}^m)|^2 \,.
\label{Eq:mag_eff}    
   \end{equation}

Equation \eqref{eq:Pm} shows several interesting aspects of magnetic power absorption in a piezoelectric--magnetostrictive bilayer resonator. First, the magnetic power absorption increases quadratically with the applied electric field, \emph{i.e.} $P_\mathrm{m}\propto E^2$. Hence, further increasing the field results in much stronger magnetization dynamics until the regime of nonlinear magnetization dynamics is reached. Moreover, the power absorption also depends quadratically on the piezoelectric constant, \emph{i.e.} $P_\mathrm{m}\propto e^2$. By contrast, the influence of magnetic material parameters, \emph{i.e.} $B$, $M_S$ and $\alpha$, is less evident since these parameters also indirectly influence the strain amplitude. In section \ref{sec:use_case}, we discuss a specific case study with particular focus on the influence of magnetic material parameters on the power absorption, both in the mechanical and magnetic subsystems. Finally, from a geometrical perspective, the power absorption strongly depends on the (relative) thickness of both the magnetic and piezoelectric layers. However, analogous to the effect of the magnetic material parameters, the thicknesses indirectly also influence the strain magnitude, and therefore, no simple general relation with magnetic power absorption can be formulated. In section \ref{sec:use_case}, this will also be studied in more detail.

The maximum power transfer into the magnetic domain is achieved when the elastic resonance coincides with the ferromagnetic resonance, \emph{i.e.}~when both $c^m_i(\omega)$ and $S^m_k(\omega)$ peak at the same frequency. However, it is difficult to tune the elastic resonance frequency without changing the dimensions of the structure. By contrast, the magnetic resonance frequency can be easily tuned by applying an external magnetic field $H_0$. In the following, we derive an approximate expression for the external field to achieve concurring resonances of both the elastic and magnetic systems. This expression can \emph{e.g.} be used to devise approximate optimal conditions for experimental realizations.

Without losses, the strain denominator $D(\omega)$ in Eq.~\eqref{Eq:S_denom} is zero at resonance, leading to an infinite strain amplitude. However, when losses are considered, the denominator becomes a complex number. At resonance, the real part of the denominator becomes zero and the strain amplitude is limited by the finite imaginary part. Hence, to find the resonance frequencies, the real part of the denominator has to be set equal to zero and the resulting equation must be solved for the frequency, \emph{i.e.} one needs to solve $D_r(\omega) = 0$ for $\omega$. However, for a bilayer system, the solution of this equation is rather cumbersome and further simplifications need to be made. Typically, the imaginary part of the complex stiffness and wavenumber is much smaller than the real part, and therefore, its influence on the resonance frequency can be neglected. Hence, we can approximate $c^m_i = c^p_i = 0$. As a result, the resonance condition becomes
\begin{equation}
     c^m_r k^m_r \cos(k^p_rd) \sin (k^m_r t)+c^p_r k^p_r \sin (k^p_rd) \cos (k^m_r t) =0\,.
\end{equation}

For a thin magnetostrictive layer, \emph{i.e.} for $t\ll d$, the solution of this equation, \emph{i.e.} the elastic resonance frequency, can be approximated by 
\begin{equation}
    \nu_\mathrm{res}^{el} \approx \frac{n v^p_r}{2(d+t)}
\end{equation}
with $n$ the mode number. This corresponds to the mechanical resonance frequency of a single piezoelectric layer of thickness $d+t$. 

Maximal power transfer can be obtained by tuning the ferromagnetic resonance frequency via the external magnetic field to coincide with the mechanical resonance frequency. The condition for concurrent resonance is
\begin{equation}
\label{eq:con}
    \nu_\mathrm{res}^{el}=\nu_\mathrm{res}^m \Rightarrow \frac{n v^p_r}{2(d+t)} = \frac{\sqrt{\omega_0(\omega_0+\omega_M)}}{2\pi}\,.
\end{equation}
The (real) magnetic field that leads to a ferromagnetic resonance frequency satisfying Eq.~\eqref{eq:con} is given by
\begin{equation}
\label{eq:H0}
    H_0 = \frac{-\omega_M + \sqrt{\omega_M^2+4\left(\frac{n\pi v^p_r}{d+t}\right)^2}}{2\mu\gamma}\,.
\end{equation}

\noindent Hence, from this expression, it can be seen that bilayers consisting of magnetostrictive materials with larger saturation magnetization and piezoelectric materials with larger group velocities require a larger external applied magnetic field to obtain concurrent mechanical and ferromagnetic resonances. 

\section{Elastic power loss}

In the previous section, we have derived expressions for the power transfer due to excitation of magnetization dynamics inside the magnetostrictive layer induced by the dynamic displacement. Below, we address the viscoelastic power loss inside the resonator. The generic expression for the elastic power density is
\begin{equation}
    p_\mathrm{el} =  \frac{i\omega}{2} \sum_{i,j,k,l} S_{ij}\left(c_{ijkl}S_{kl}\right)^* \,.
\end{equation}
In the geometry discussed in this work, only the shear strain $S_{xz}=S_{zx}$ contributes to the mechanical energy in the resonator. For systems with tetragonal or hexagonal crystal symmetry, the power density contained in the shear strain becomes
\begin{equation}
    p_\mathrm{el}(z) =  i\omega \left(c^{p*}_t |S^p(z)|^2 + c^m_0 |S^m(z)|^2 \right) \,.
\end{equation}
Similar to the magnetic power, the power per unit area is obtained by integration over the resonator thickness, leading to
\begin{align}
    \cal{P}_\mathrm{el} &= i\omega \left(c^{p*}_t \int\limits_{0}^{d}|S^p(z)|^2 dz+ c^m_0 \int\limits_{-t}^{0}|S^m(z)|^2 dz \right)\\
 \label{approx}    &\approx  i\omega c^{p*}_t  \int\limits_{0}^{d}|S^p(z)|^2 dz\\
     & =  i\omega (c^p_r - ic^p_i) e^2 E^2  \int\limits_{0}^{d}|S^p_k(z)|^2 dz \,.
\end{align}
\noindent Here, the approximation in Eq.~\eqref{approx} is valid when the magnetostrictive layer is much thinner than the piezoelectric layer, as it is assumed throughout the paper. The real part of this expression represents the elastic power loss due to viscoelastic effects and is given by
\begin{equation}
\label{eq:Pel}
P_\mathrm{el} = \omega E^2 \frac{c^p_re^2}{Q} \int\limits_{0}^{d}|S^p_k(z)|^2 dz \,.
\end{equation}
As for the magnetic power absorption, the elastic power loss depends on the square of the applied electric field and the piezoelectric constant, \emph{i.e.} $P_\mathrm{el}\propto e^2E^2$. Since other parameters also influence the strain magnitude, again, no simple relations with the power loss can be established. In this case, numerical calculations are required, \emph{e.g.}~to better understand the influence of material parameters or resonator thickness.

\section{Magnetoelectric transduction efficiency\label{sec:eta}}

When the resonator is used as a magnetoelectric transducer to excite magnetization dynamics from electric signals, the power transfer efficiency in the magnetic domain is a key performance parameter. The total power delivered to the system can be partitioned into three main parts: the reactive power, the elastic power loss, and the magnetic power absorption. The reactive part corresponds to the power that resonantly oscillates inside the transducer. This power is not lost and needs to be supplied only during the transient regime of the resonator before a steady state is reached. By contrast, the elastic and magnetic power losses can be used to describe a transducer efficiency. In an ideal magnetoelectric transducer, losses occur mainly in the magnetic subsystem, while elastic losses can be considered as parasitic and need to be minimized. Note that the elastic losses mainly occur in the piezoelectric medium as its volume is much larger than the magnetostrictive volume.

By comparing mechanical and magnetic losses, the magnetic transduction efficiency of the magnetoelectric resonator can be determined

\begin{equation}
\label{eq:eff}
\eta = \frac{P_\mathrm{m}}{P_\mathrm{m} + P_\mathrm{el}}\, , 
\end{equation}

\noindent with $P_\mathrm{m}$ and $P_\mathrm{el}$ given by Eqs.~\eqref{eq:Pm} and \eqref{eq:Pel}, respectively.

The efficiency represents the fraction of the active power in the resonator that is converted into magnetic excitations. Note that this efficiency is different from the \emph{absolute} power dissipated in the magnetic subsystem. Certain parameters affect both the efficiency as well as the total magnetically dissipated power and therefore, the efficiency can increase in certain cases whereas the total magnetically dissipated power decreases. This will be discussed further in the case study below.

Equations \eqref{eq:Pm} and \eqref{eq:Pel} indicate that the efficiency is independent of the resonator area, the piezoelectric constant $e$, and the applied electric field $E$. This leads to several interesting conclusions. First, in our model, the efficiency is unaffected by reducing the transducer area, which indicates a favorable scaling behavior of the resonator as a transducer. Furthermore, a reduction of the applied electric field or the voltage does not affect the transducer efficiency but impacts only the total dissipated power. More surprisingly, also the piezoelectric constant does not affect the transducer efficiency. This suggest that the viscoelastic losses and the mechanical $Q$-factor of the resonator are key parameters for the transducer efficiency, while the piezoelectric coupling constant only affects the \emph{total} power loss. 

\section{Case study: S\lowercase{c}A\lowercase{l}N resonators with different magnetostrictive layers}
\label{sec:use_case}

As an illustration of the above model and to gain more quantitative insight in the expected behavior of magnetoelectric resonators, we study in this section a concrete system based on hexagonal ScAlN as the base piezoelectric material.\cite{fichtner_alscn:_2019,petrich_investigation_2019,wang_film_2020} The ScAlN material parameters were obtained from the literature:\cite{Caro_2015} piezoelectric coupling constant $e=0.3$~C/m$^2$, mass density $\rho^p=3.5$~g/cm$^3$, and stiffness constant $c_{55}^p=c^p_r=100$~GPa. The piezoelectric layer thickness was set to $d=200$~nm. The $Q$-factor corresponding to the individual piezoelectric layer was assumed to be $Q=1000$, unless specified otherwise. It was then used to determine the imaginary part of the piezoelectric stiffness via $c^p_i=c^p_r/Q$.\cite{Park2019} As magnetostrictive layers, we have studied three materials: CoFeB, Ni, and Terfenol-D (Tb$_{0.3}$Dy$_{0.7}$Fe$_2$). The material parameters of the magnetostrictive materials have also been obtained from the literature and are summarized in Tab.~\ref{table:material_params}. 

\begin{table}[tb]
\begin{tabular}{l c c c}
\hline\hline
Material         &  CoFeB & Ni & Terfenol-D \\ \hline
$\rho$ (g/cm$^3$)  &    8   &  8.9  & 9.3    \\ 
$c^m_0$ (GPa)        &    70   &   74 & 38    \\ 
$B$ (MJ/m$^3$)     &     7  &  18  &  118   \\ 
$M_S$ (kA/m)        &     1000  & 370   & 700    \\ 
$\alpha$ ($\times 10^{-3}$)          &    4   &  10  &  60   \\ \hline\hline
\end{tabular}
\caption{Material parameters for CoFeB,\cite{Peng2016} Ni,\cite{Dreher2012} and Terfenol-D (Tb$_{0.3}$Dy$_{0.7}$Fe$_2$):\cite{Colussi2016,Gopman2016,Wang2017} mass density $\rho$, bulk stiffness $c^m_0$, magnetoelastic coupling constant $B$, saturation magnetization $M_S$, and Gilbert damping constant $\alpha$.}
\label{table:material_params}
\end{table}

\begin{figure}[p]
	\includegraphics[width=8.5cm]{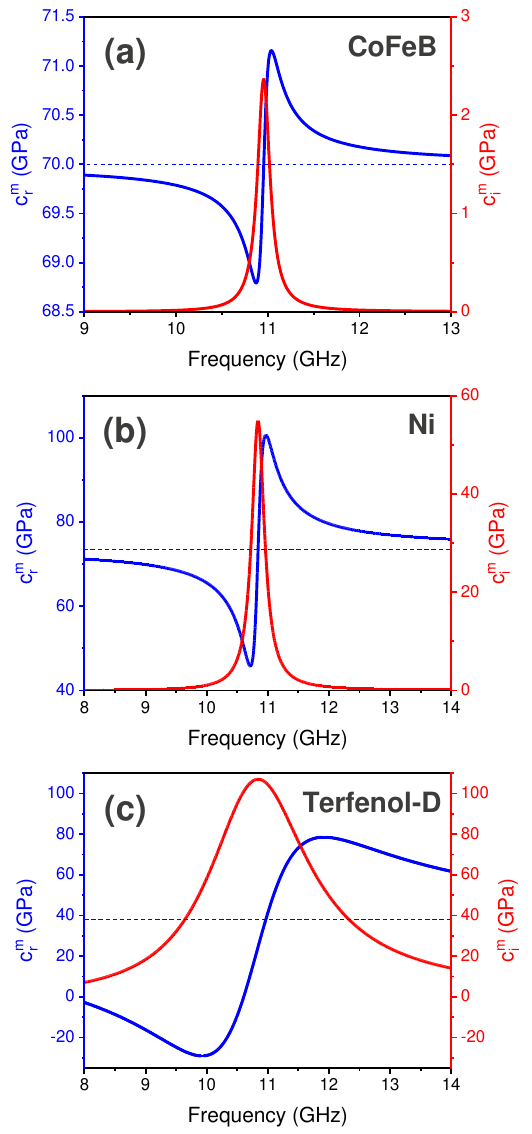}
	\caption{Real and imaginary parts of the effective stiffness of (a) CoFeB, (b) Ni, and (c) Terfenol-D as a function of frequency near ferromagnetic resonance. The dotted lines correspond to the CoFeB, Ni, and Terfenol-D stiffnesses in absence of magnetoelastic coupling. The external applied field was $\mu_0H = $111.75 mT, $\mu_0H = $218.97 mT and $\mu_0H = $145.70 mT for CoFeB, Ni, and Terfenol-D, respectively}
	\label{fig:cm}
\end{figure}

The frequency dependence of the (complex) effective stiffness $c^m$ of CoFeB, Ni, and Terfenol-D near ferromagnetic resonance are plotted in Fig.~\ref{fig:cm}. In the graph, the blue dotted lines correspond to stiffnesses for zero magnetoelastic coupling whereas the solid blue and red lines indicate the real and imaginary parts of the effective stiffness, respectively, when the magnetoelastic coupling is active. The data indicate that the magnetization dynamics render the real part of the stiffness frequency dependent and induce an additional imaginary component. The imaginary component accounts for the losses in the magnetic system. For CoFeB (Fig.~\ref{fig:cm}(a)), the maximal magnitude of the imaginary part is comparable to the frequency modulation of the real part, about $30\times$ lower than the unperturbed stiffness $c^m_0$. The frequency dependence of the effective stiffness corresponds to that of the dynamic magnetic susceptibility since the effective stiffness is proportional to the susceptibility. Hence, the frequency band, over which the magnetic effect is present, strongly depends on the width of the (ferromagnetic or spin wave) resonance of the magnetic system, which in turn depends on the magnetic damping parameter $\alpha$. We note that the magnetic resonance width, and thus also the width of the frequency dependence of $c^m_i$, is independent of the elastic resonance width, which is determined by the mechanical resonator $Q$-factor. Typically, the magnetic resonance is (much) broader than the mechanical resonance (as in this case study), although exceptions can occur for low-$Q$ FBARs. 

For comparison, the frequency dependence of the effective stiffness of Ni and Terfenol-D (Tb$_{0.3}$Dy$_{0.7}$Fe$_2$) are plotted in Figs.~\ref{fig:cm}(b) and \ref{fig:cm}(c), respectively. Due to the larger magnetostriction of Ni and Terfenol-D, the frequency dependent part of the effective stiffness $c^m$ is now comparable or even larger than the intrinsic elastic stiffness in absence of magnetostriction. Note that this even can result in negative effective stiffnesses below the resonance frequency for Terfenol-D. Moreover, the frequency band, over which the effective stiffness is affected by magnetostriction, is much larger than for CoFeB due to the much larger magnetic damping of Ni and Terfenol-D. Consequently, for Terfenol-D the mechanical properties are dominated by magnetic effects near ferromagnetic resonance whereas the magnetic effects can be seen more as a perturbation in case of CoFeB.

Other material-dependent factors that affect the effective stiffness are the saturation magnetization $M_S$ and the Gilbert damping (see Eq.~\eqref{eq:cires_m}). Ni possesses a low saturation magnetization (see Tab.~\ref{table:material_params}), which enhances the magnetoelastic coupling. By contrast, Terfenol-D is characterized by strong Gilbert damping $\alpha$, which reduces the magnetoelastic coupling. As a result, the imaginary parts of the effective stiffness of Ni and Terfenol-D are comparable despite the much larger magnetoelastic coupling coefficient $B$ of Terfenol-D. For CoFeB, the low $\alpha$ cannot compensate for the comparatively low $B$, especially since the saturation magnetization is large.

As mentioned above, the magnetic resonance frequency can be tuned by an external magnetic field $H_0$ (see Eq.~\eqref{eq:H0} for an approximate expression) to match the mechanical frequency of the resonator, leading to strong elastic--magnetic coupling conditions. Below, we will discuss power losses in both the mechanical and magnetic subsystems and the transduction efficiency under such resonance-matched conditions. We first discuss effects of device dimensions and compare the performance of the different magnetostrictive materials: CoFeB, Ni, and Terfenol-D. Subsequently, we discuss in more detail the effect of materials parameters, specifically the magnetoelastic coupling constant and the magnetic Gilbert damping. Finally, in the last subsection, we address the impact of mechanical resonator losses by assessing the effect of the mechanical $Q$-factor on the magnetic excitation efficiency.

\subsection{Magnetostrictive material dependence and figure of merit}

\begin{figure}[p]
	\includegraphics[width=8cm]{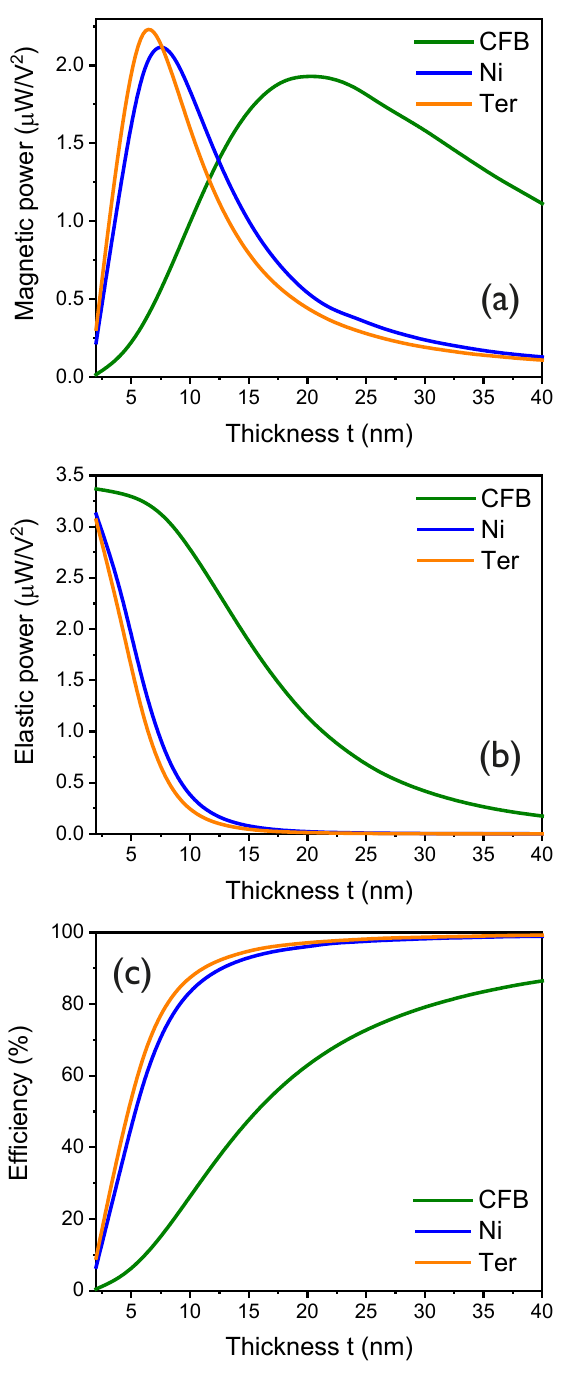}
	\caption{(a) Magnetic power loss, (b) elastic power loss, and (c) magnetic transduction efficiency $\eta$ as a function of magnetostrictive layer thickness $t$ using CoFeB (CFB), Ni, and Terfenol-D (Ter) material parameters. In all cases, elastic and ferromagnetic resonances were matched by adjusting the external magnetic field.}
	\label{fig:t_mats}
\end{figure}

The effective stiffness $c^m$ of the magnetostrictive layer depends solely on material parameters and is not influenced by device dimensions. Nevertheless, the dimensional parameters influence the strain amplitude and the volume integration and thus modify power transfers and transducer efficiency. To illustrate this dimensional influence, the magnetic and elastic power loss per area and squared electric field, \emph{i.e.} $P_m/E^2$ and $P_\mathrm{el}/E^2$, have been calculated for the three considered materials (CoFeB, Ni, and Terfenol-D) as a function of magnetostrictive layer thickness $t$. The results are shown in Figs.~\ref{fig:t_mats}(a) and \ref{fig:t_mats}(b), respectively. Note that for every thickness, the external magnetic field has been adjusted so that the magnetic resonance frequency matches the elastic resonance frequency, resulting in maximal power absorption.

The data show that, for all materials, the elastic power gradually decreases with magnetostrictive layer thickness, whereas the magnetic power possesses a maximum at a finite thickness. This general behavior can be explained by the variation of the strain amplitude near resonance since strain plays a key role in both the magnetic as well as the elastic power loss. At resonance, the strain amplitude peaks when the real part of the strain denominator becomes zero, \emph{i.e.} $D_r(\omega_r) = 0$, and it becomes limited only by the imaginary part of the strain denominator, which reflects the losses in the system, \emph{i.e.} $S(\omega_r)\propto 1/D_i$. In a magnetoelastic resonator, both elastic and magnetic losses contribute to the imaginary part of the strain denominator. When magnetic losses are low, the imaginary part of the strain denominator as well as the strain amplitude are determined by elastic losses. This means that the strain amplitude is mainly determined by the properties of the piezoelectric layer and not by the magnetic material properties. By contrast, for high magnetic losses, the strain denominator and the strain amplitude are strongly influenced by the magnetic layer properties.

For thin magnetostrictive layers, the magnetic power loss in the resonator is low compared to the elastic power loss due to a small magnetic volume. This means that the magnetic contribution to the strain denominator can be neglected in this case, resulting in large strain amplitudes and large elastic power absorption. For $t=0$, the elastic power dissipation reaches a maximum as the mechanical resonance is not perturbed by an additional magnetostrictive layer. When the magnetic layer thickness increases, the magnetostrictive (magnetic) volume increases and thus also the magnetic power absorption. When the magnetic power reaches a similar magnitude as the elastic power loss, the magnetic contribution to the strain denominator cannot be neglected anymore, which results in even stronger decreasing strain amplitude and therefore in decreasing elastic power. At one point, this decreasing strain amplitude also affects the magnetic power and offsets the effect of an increasing magnetic volume, leading to a peak in the magnetic power at finite $t$. 

A comparison of the results for the three materials suggests that the peak occurs at smaller thicknesses for stronger magnetostriction. The difference between the materials can be understood from the imaginary part of the effective stiffness $c^m_i$ that determines how much power can be absorbed for a given strain amplitude. Fig.~\ref{fig:cm} shows that the peak values of $c^m_i$ are much larger for Terfenol-D and Ni than for CoFeB because of their higher magnetoelastic coupling constant $B$. As a result, Terfenol-D and Ni absorb more power for a given strain amplitude and thus require less magnetic volume to reach a magnetic power absorption that is equal to the elastic one. 

Figure \ref{fig:t_mats}(c) depicts the magnetic transduction efficiency $\eta$, defined in Eq.~\eqref{eq:eff}, as a function of magnetostrictive layer thickness $t$. The efficiency is a key metric when magnetoelectric resonators are used as magnetic transducers. For all materials, the efficiency increases with magnetoelectric thickness $t$ as the magnetically dissipated power increases with $t$, whereas the elastic power decreases. It is noteworthy that for the chosen parameters (in particular $Q=1000$), very high magnetic transduction efficiencies close to 100\% can be achieved for Terfenol-D but also for Ni. Values for CoFeB are somewhat lower but still reach values of 80\% for layers as thin as 40 nm. This confirms expectations that magnetoelectric transducers can be highly efficient to excite ferromagnetic resonance (or spin waves), and that magnetic losses can exceed elastic ones for the parameters considered here.

Based on the above discussion, we can define a figure of merit (FoM) for the magnetostrictive materials in a magnetoelectric resonator. According to Eq.~\eqref{eq:ci_m}, the effective stiffness depends on the magnetic material parameters and the frequency. Since in this approach, the magnetostrictive layer is considered to be thin, it does not affect much the mechanical resonance frequency. As a result, the frequency factor in $c_i^m$ can be omitted when comparing different magnetostrictive materials and the FoM can be defined as

\begin{equation}
\label{eq:FoM}
\mathrm{FoM} = \frac{B^2H_0}{M_S\alpha(2H_0+M_S)}\,.
\end{equation} 

Materials with a higher FoM dissipate more magnetic power and reach higher efficiencies $\eta$ for a specific magnetic layer thickness $t$. To illustrate this, the external magnetic field for matched ferromagnetic and elastic resonances, the resonance frequency, the magnetic and elastic power absorption, the magnetic transduction efficiency, and the FoM have been calculated for 20 nm thick CoFeB, Ni, and Terfenol-D layers. The results are summarized in Tab.~\ref{table:efficiency}. From this table, it can be seen that the resonance frequency for the three materials is roughly the same and the magnetic transduction efficiency correlates positively with the FoM. The highest efficiency and FoM is obtained for Terfenol-D because of its high magnetoelastic coupling constant. 

The FoM gives insight into the efficiency but does not explain the absolute power absorption. To better understand the absolute power absorption, the influence of the material parameters needs to be further investigated. This will be the topic of the following sections. Between the different magnetostrictive materials (see Tab.~\ref{table:material_params}), the mass density $\rho^m$, the stiffness $c^m$, and the saturation magnetization $M_S$ only differ by a rather small factor $< 3$ and thus have minor influence on relative power absorption of the different materials. On the other hand, the magnetoelastic coupling constant $B$ and the Gilbert damping constant $\alpha$ can differ by more than an order of amplitude and are thus expected to have a considerable influence on power absorption. Hence, we will focus on these parameters in the following.

Beyond the influence of material parameters, we note that also the total resonator thickness (dominated by the piezoelectric thickness) plays a role. A thinner resonator leads to a higher mechanical resonance frequency as well as to, due to frequency matching, higher $H_0$. This strongly affects the magnetic power absorption (see Eq.~\eqref{Eq:mag_eff}) and also the FoM in Eq.~\eqref{eq:FoM}. 

\begin{table}[]
\begin{tabular}{l c c c}
\hline\hline
Material & CoFeB & Ni & Terfenol-D \\ \hline
$\mu_0H$ (mT)           &      111.8          &      219.0       &     145.7         \\ 
$\nu_\mathrm{res}$ (GHz)      &        10.96        &       10.84      &        10.83      \\ 
$P_m/E^2$ ($\mu$W/V$^2$)    &         1.94       &       0.53      &       0.44       \\ 
$P_{el}/E^2$ ($\mu$W/V$^2$) &         0.57       &        0.010     &        0.0065      \\ 
$\eta$ (\%)      &         77       &       98      &         99     \\ 
FoM  (10$^6$ N$^2$/Am$^3$)  &       1.4         &       27      &       69       \\ \hline\hline
\end{tabular}
\caption{External applied field $\mu_0H$, resonance frequency $\nu_\mathrm{res}$, magnetic and elastic power absorption, magnetic transduction efficiency $\eta$, and FoM for CoFeB, Nickel and Terfenol-D respectively. Piezoelectric and magnetostrictive layer thicknesses are 200 nm and 20 nm respectively}
\label{table:efficiency}
\end{table}

\subsection{Impact of the magnetoelastic coupling constant}

In this section, we investigate the effect of the magnetoelastic coupling constant $B$ on the power absorption in the resonator. In this and the following section, the magnetic layer thickness has been set to 20 nm. To isolate the effect of $B$, all other materials parameters have been fixed to those of CoFeB. We note again that in all cases, $Q = 1000$.

\begin{figure}[tb]
	\includegraphics[width=8.5cm]{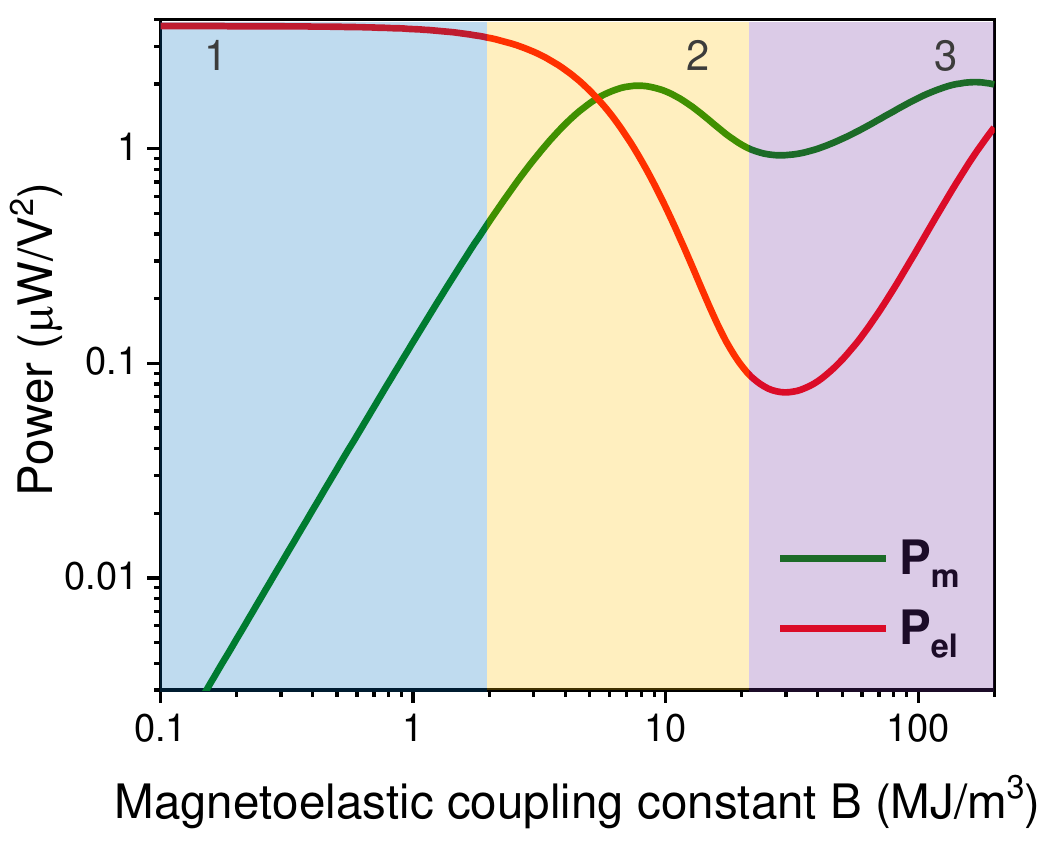}
	\caption{Magnetic ($P_\mathrm{m}$) and elastic ($P_\mathrm{el}$) power absorption as a function of magnetoelastic coupling constant $B$. All other material parameters were fixed to those of CoFeB, and $Q = 1000$. Three regimes can be identified (differently shaded regions), as discussed in the text.}
	\label{fig:B_reg}
\end{figure}

Figure~\ref{fig:B_reg} shows the magnetic and elastic power absorption under these conditions for varying $B$. In the figure, different regimes can be identified, which are indicated by different background shading. At low $B \lesssim 2$ MJ/m$^3$ (blue shaded region), the mechanical power loss dominates and is nearly independent of $B$. By contrast, the magnetic power absorption increases quadratically with $B$ via its dependence on $c^m_i$. Since the elastic losses are dominant, they limit the strain denominator at resonance. Consequently, the strain amplitude is determined by the piezoelectric properties and is independent of the magnetic system. As a result, increasing $B$ does not affect the strain amplitude significantly. Hence, the total power absorption increases (weakly) with $B$ in this regime because the large and constant elastic power loss dominates the increasing magnetic power loss.

At higher $B$ values between 2 and about 20 MJ/m$^3$, the magnetic and elastic power losses reach similar magnitudes (yellow shaded region in Fig.~\ref{fig:B_reg}). In this regime, the magnetic contribution to the imaginary part of the strain denominator cannot be neglected anymore, which results in decreasing strain amplitudes at resonance for increasing $B$. The decreasing strain results in decreasing elastic power and also slows down the magnetic power absorption increase until a maximum is reached. Beyond this maximum, the decreasing strain has a stronger effect than the increasing magnetoelastic coupling, resulting in decreasing magnetic power, together with a strongly decreasing elastic power loss. Hence, in this regime, the total power absorption of the resonator \emph{decreases} with increasing $B$.

In the third regime for $B \gtrsim 20$ MJ/m$^3$ (purple shaded region in Fig.~\ref{fig:B_reg}), both the magnetic and elastic power increase again after reaching a minimum. For such large $B$, the effective stiffness $c^m_0$ is dominated by the magnetic contribution $c^m_B$. Under these conditions, a strong acoustic impedance mismatch between the piezoelectric and magnetostrictive layers exists and there is significant acoustic reflection at their interface. Consequently, the acoustic dynamics in the two layers become increasingly decoupled, which leads to strongly modified elastodynamics both in the piezoelectric as well as the magnetostrictive layer. Due to the decoupling, the strain in the decoupled piezoelectric layer increases again rapidly with $B$, whereas the strain in the decoupled magnetostrictive layer shows a weaker increase, as shown in Fig.~\ref{fig:B_reg}. 

\begin{figure}[p]
	\includegraphics[width=7.8cm]{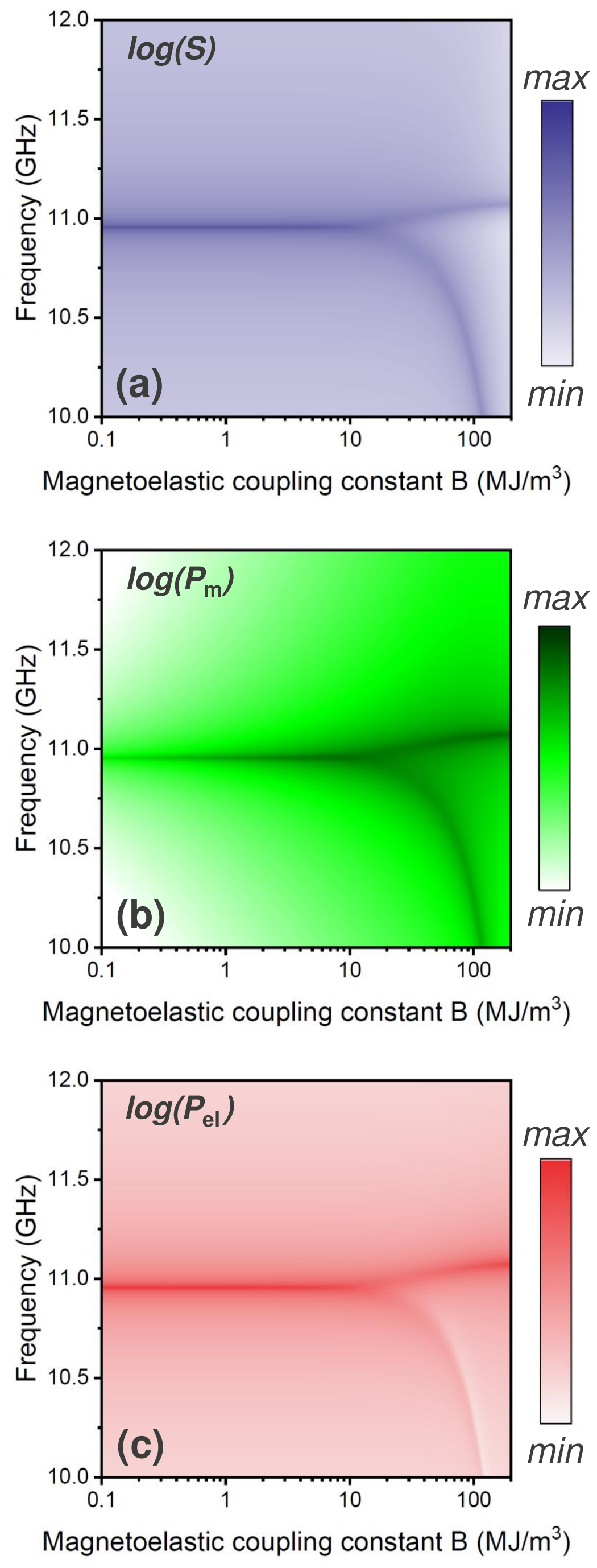}
	\caption{(a) Strain, (b) magnetic power loss $P_\mathrm{m}$, and (c) elastic power loss $P_\mathrm{el}$ \emph{vs.}~frequency and magnetoelastic coupling constant. All other material parameters were those of CoFeB, and $Q = 1000$. The single resonance frequency splits into two at large $B$.}
	\label{fig:B_map}
\end{figure}

In addition to the decoupling between the layers, high $B$ values also lead to a splitting of the resonance frequency, as shown by Fig.~\ref{fig:B_map}. This splitting originates from the magnetic contribution to the real part of the effective stiffness $c_r^m$, which becomes non-negligible at large $B$. As shown in Fig.~\ref{fig:cm}, the real part $c_r^m$ has two antisymmetric extrema, one on each side of the ferromagnetic resonance frequency. These two peaks result in two distinct zeros in the real part of the strain denominator. Since the zeros in $D_r$ determine the peak response of the resonator, this result in two split resonance frequencies. This splitting leads to two frequency branches of both the strain amplitude, as well as the magnetic ($P_\mathrm{m}$) and elastic ($P_\mathrm{el}$) power losses, as illustrated in Fig.~\ref{fig:B_map}. The splitting becomes visible at around $B\approx10$--20 MJ/m$^3$. We note that the intrinsic CoFeB magnetoelastic coupling constant is $B\approx7$~MJ/m$^3$, for which the splitting is not yet visible. Hence, intrinsic CoFeB leads to a regime 2 behavior, as indicated by Fig. \ref{fig:B_reg}. 

\begin{figure}[p]
\label{fig:B}
	\includegraphics[width=8.5cm]{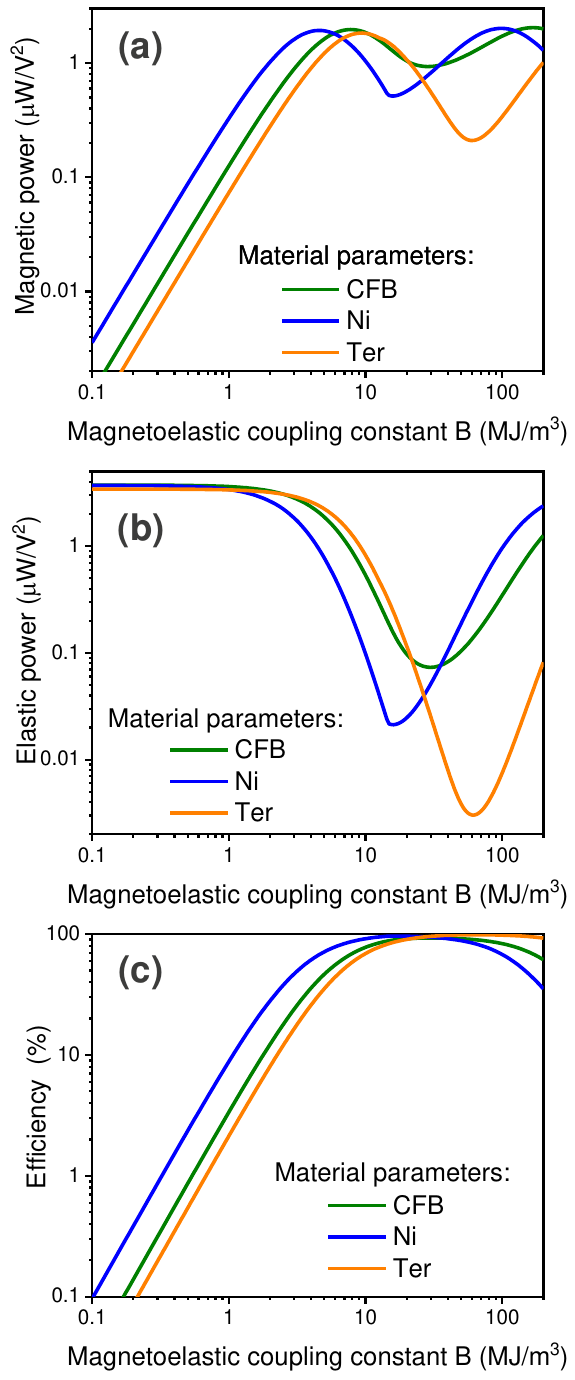}
	\caption{(a) Magnetic power loss $P_\mathrm{m}$, (b) elastic power loss $P_\mathrm{el}$, and (c) magnetic transduction efficiency $\eta$ as a function of the magnetoelastic coupling constant $B$ for materials with CoFeB (CFB), Ni, and Terfenol-D (Ter) parameters.}
	\label{fig:B_graph}
\end{figure}

In the above discussion, we have assumed CoFeB material parameters throughout, except for $B$, which was varied. To see whether the above behavior is universal or depends critically on material properties, we have also calculated the dependence of magnetic ($P_\mathrm{m}$) and elastic ($P_\mathrm{el}$) power losses as well as of the magnetic transduction efficiency $\eta$ on $B$ for Ni and Terfenol-D material properties. The results are shown in Fig.~\ref{fig:B_graph}. The data show similar overall trends for the three sets of material prarameters, \emph{i.e.} a low-$B$ regime with dominant elastic losses, an intermediate regime with similar elastic and magnetic losses, and a high-$B$ regime with dominant magnetic losses and the mechanical decoupling of the two layers. Different material parameters mainly lead to two effects: (i) a moderate shift of the $B$ onset values of the different regimes, with Terfenol-D requiring slightly higher $B$ to reach regimes 2 and 3; and (ii) an increased depth of the minimum of magnetic and elastic power loss at the onset of regime 3 for Terfenol-D. This indicates that the above dependence of magnetic and elastic power losses on $B$ is rather generic and does not depend critically on material parameters. It is worth noting that the maximum magnetic transduction efficiency $\eta$ is reached for similar $B$ values for all three sets of material parameters. However, some small differences exist and Ni material parameters (especially the low $M_S$) can compensate (partially) for the lower $B$ with respect to Terfenol-D (see also Tab~\ref{table:efficiency}). 

\subsection{Impact of the magnetic Gilbert damping}

\begin{figure}[tb]
	\includegraphics[width=8.5cm]{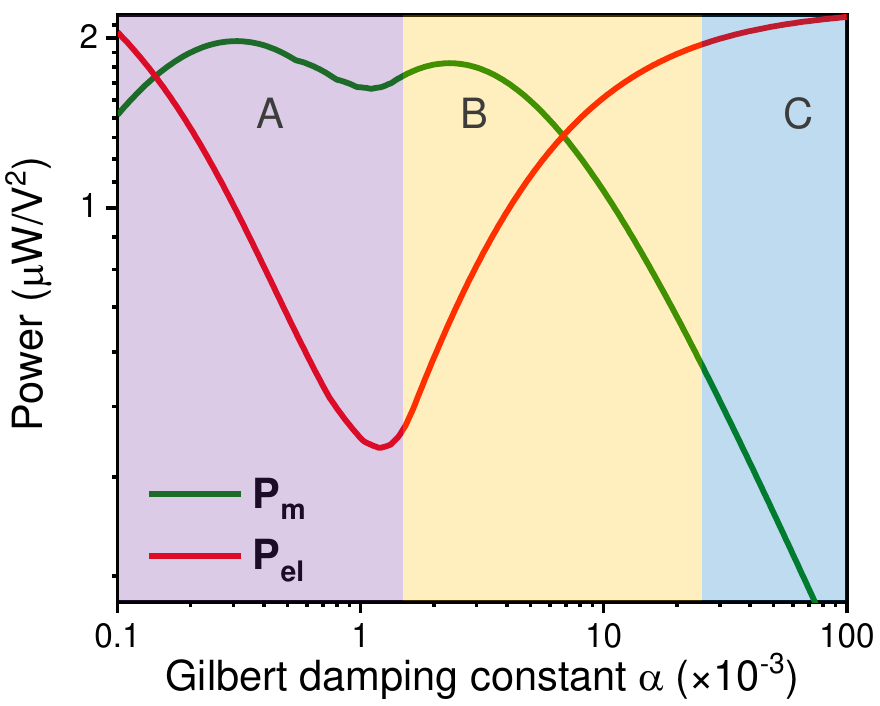}
	\caption{Magnetic ($P_\mathrm{m}$) and elastic ($P_\mathrm{el}$) power loss for a material with CoFeB parameters and varying Gilbert damping constant $\alpha$. Three regimes are identified that are color encoded.}
	\label{fig:a}
\end{figure}

We now turn to the impact of the Gilbert damping constant $\alpha$ of the magnetostrictive material. As in beginning of the previous section, we fix all material parameters to those of CoFeB, except for $\alpha$, which is varied. In general, $\alpha$ has the opposite effect as that of the magnetoelastic coupling constant $B$, since the imaginary part of the stiffness constant is inversely proportional to $\alpha$, \emph{i.e.} $P_m\propto c^m_i \propto B^2/\alpha$. As a consequence, the dependences of the magnetic ($P_\mathrm{m}$) and elastic ($P_\mathrm{el}$) power losses on $\alpha$ in Fig.~\ref{fig:a} show a ``mirrored'' behavior of that \emph{vs.} $B$ in Fig.~\ref{fig:B_reg}. Hence, three regimes can again be identified (marked by different shadings in Fig.~\ref{fig:a}) with the same mechanisms as discussed in the previous section. This means that the low-$\alpha$ regime A corresponds to the high $B$ regime 3, in which the two layers are mechanically decoupled from each other, and the resonance frequency is split. Increasing $\alpha$ results in regime B, in which the magnetic power loss is of similar magnitude as the elastic power loss (equivalent to regime 2 for intermediate $B$). At high $\alpha$, the effective stiffness is low, resulting in low magnetic power loss. As a result, the elastic power loss becomes dominant and the strain amplitude becomes constant (regime C, equivalent to regime 1 in Fig. \ref{fig:B_reg}).

In addition to the power loss amplitude, the magnetic damping also influences the resonance width of the system. In regime C, where the elastic losses are dominant, the strain denominator and thus the resonance width is determined by the piezoelectric layer properties, in particular by the mechanical $Q$-factor of the resonator in absence of magnetoelastic coupling. However, in regimes A and B, the magnetic contribution cannot be neglected, and the resonance width converges to the resonance width of the magnetic system, determined by the magnetic damping constant.

\begin{figure}[p]
	\includegraphics[width=8cm]{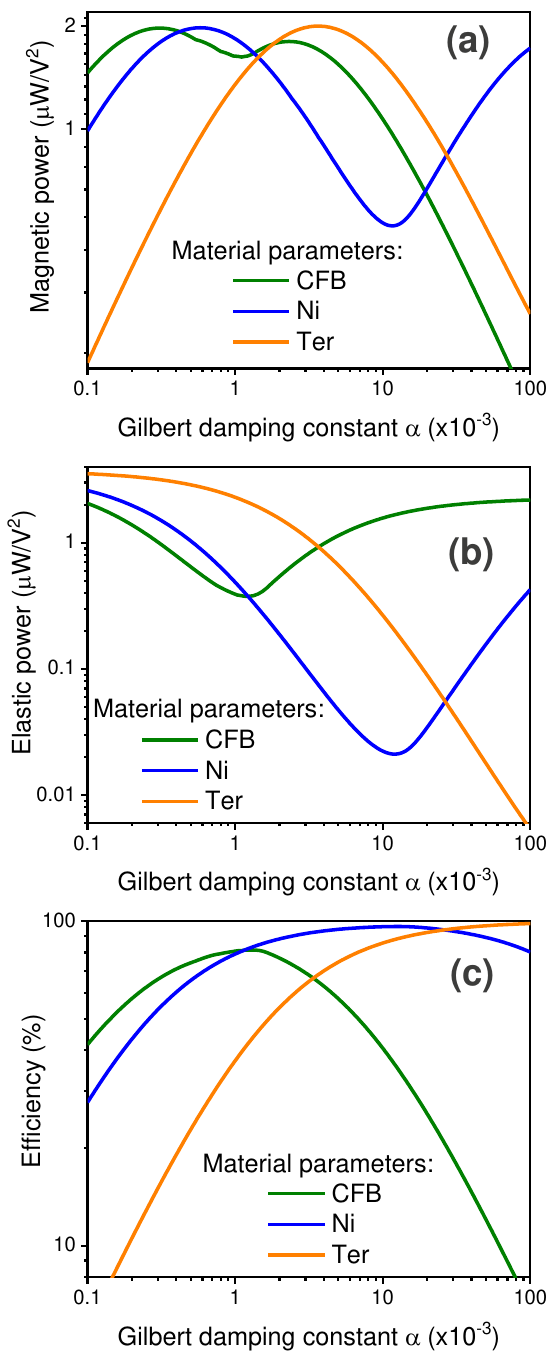}
	\caption{(a) Magnetic power loss $P_\mathrm{m}$, (b) elastic power loss $P_\mathrm{el}$, and (c) magnetic transduction efficiency $\eta$ as a function of the Gilbert damping constant $\alpha$ for materials with CoFeB (CFB), Ni, and Terfenol-D (Ter) parameters.}
	\label{fig:as}
\end{figure}

As in the previous section, calculations for different sets of material parameters (with the exception of $\alpha$, which is varied) corresponding to CoFeB, Ni, and Terfenol-D indicate that this behavior is generic (see Fig.~\ref{fig:as}). Both the magnetic power loss, the elastic power loss, and the magnetic transduction efficiency show similar trends for materials with CoFeB, Ni, and Terfenol-D parameters with varying $\alpha$. Analogous to the dependence on the magnetoelastic coupling constant $B$, the efficiency shows a maximum as a function of $\alpha$ when the elastic power loss reaches a minimum. We note that the actual power loss and efficiency performance of the different materials depend on the actual values of $\alpha$ and $B$. Results are summarized in Tab.~\ref{table:efficiency}.

\subsection{Impact of the mechanical $\boldsymbol{Q}$-factor of the resonator}

The previous sections discuss how magnetostrictive or magnetic material parameters affect the magnetic and elastic power absorption in the resonator as well as the magnetic transduction efficiency. It is also clear, however, that the $Q$-factor will have a major influence, as discussed before in Sec.~\ref{sec:eta}. Previously, we have fixed the $Q$-factor at a value of 1000, which is representative for high-quality FBARs. Here, we now discuss how power losses and transduction efficiency are affected by the $Q$-factor for CoFeB, Ni, and Terfenol-D material parameters. In all cases, the magnetostrictive layer was 20 nm thick and combined with a 200 nm thick ScAlN piezoelectric layer.

\begin{figure}[tb]
	\includegraphics[width=16cm]{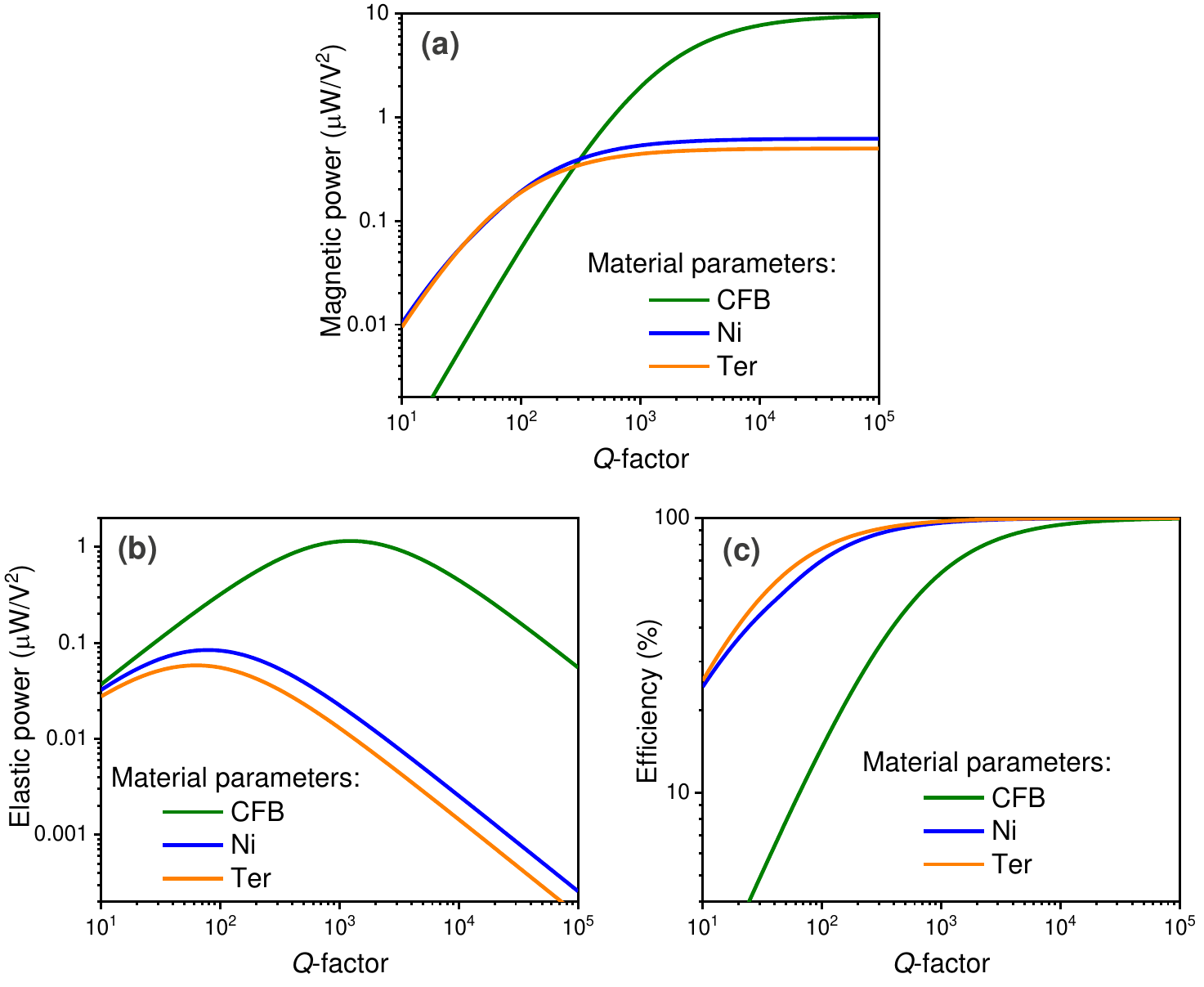}
	\caption{(a) Magnetic power loss $P_\mathrm{m}$, (b) elastic power loss $P_\mathrm{el}$, and (c) magnetic transduction efficiency $\eta$ as a function of the mechanical $Q$-factor of the resonator. Results are shown for different magnetostrictive materials CoFeB (CFB), Ni, and Terfenol-D (Ter).}
	\label{fig:Q}
\end{figure}

The results are depicted in Fig.~\ref{fig:Q}. The data indicate that the magnetic power absorption strongly increases with increasing $Q$ at low values and then saturates at high $Q$ values (Fig.~\ref{fig:Q}(a)). By contrast, the elastic power absorption first increases, reaches a maximum, and then decreases with increasing $Q$-factor (Fig.~\ref{fig:Q}(b)). This behavior can be explained by the dependence of the strain amplitude inside the device on $Q$. As intuitively expected, elastic losses in the piezoelectric dominate the resonator behavior at low $Q$. Increasing the $Q$-factor (\emph{i.e.}~reducing mechanical losses) results in decreasing imaginary part of the strain denominator and consequently increases the strain amplitude. The larger strain leads (quadratically) to an increasing magnetic power absorption, according to Eq.~\eqref{eq:Pm}, \emph{i.e.}~$P_{m}\propto S^2$. By contrast, the elastic power also grows quadratically with strain but also depends on the imaginary component of the effective stiffness, which is inversely proportional to the $Q$-factor, \emph{i.e.}~$P_{el}\propto c_i^pS^2 \propto S^2/Q$. Hence, the elastic power loss increases less strongly with $Q$ than the magnetic power loss. 

For even larger $Q$-values, the magnetic and elastic power losses become similar in magnitude. In this regime, the strain amplitude is not only determined by the $Q$-factor since the magnetic contribution to the strain denominator cannot be neglected anymore. At the highest $Q$-factors, the magnetic loss dominates the strain denominator, and the strain amplitude depends solely on the magnetic layer properties and not on the $Q$-factor anymore. As a result, in this regime, further increasing $Q$ does not influence the strain anymore, and the magnetic power loss is saturating. On the other hand, the elastic power loss still depends on the $Q$-factor via the complex stiffness and thus continues to decrease with increasing $Q$. This behavior explains the peaking elastic power loss \emph{vs.}~$Q$. As above, the behavior is generic and does not qualitatively depend on the magnetostrictive material properties. 

The resulting magnetic transduction efficiency $\eta$ is depicted as a function of $Q$ in Fig.~\ref{fig:Q}(c). Since the magnetic power absorption increases rapidly with $Q$ whereas the elastic power absorption increases more slowly or even decreases, the efficiency increases with $Q$ and subsequently saturates at unity. Note that $\eta$ can reach high values for Ni and Terfenol-D even at comparatively low $Q$-factors in the range of 100 or even below. 

\section{Conclusion}

In this work, we have derived an analytical model for the power transfer in a magnetoelectric resonator consisting of a piezoelectric--magnetostrictive bilayer. The differential equations describing the dynamics for both layers were derived and solved for free boundary conditions. The general solution was presented and utilized to derive expressions for both the magnetic and elastic power absorption inside the device as a function of frequency, device dimensions, and material parameters. Based on the power absorption expressions, the magnetic transduction efficiency was determined. The magnetic power absorption strongly depended on the complex part of the effective stiffness that captures the influence of the dynamic magnetization on the elastodynamics. Besides the effective stiffness, both the magnetic and elastic power absorption were found to strongly depend on the strain inside the device.

In the second part of the work, a case study based on an example bilayer structure (200 nm ScAlN, 20 nm CoFeB, Ni, or Terfenol-D) has been used to illustrate the influence of dimensional and material parameters on magnetic and elastic power losses and the transducer efficiency. Different regimes could be identified that depend on the ratio between the magnetic and elastic power losses. For dominant elastic losses, it was found that the piezoelectric layer parameters determine the transducer efficiency, mainly via the mechanical $Q$-factor of the resonator. By contrast, when the magnetic losses were dominant, the magnetostrictive and magnetic material parameters ($B$, $\alpha$) determined the power losses and the transducer efficiency. 

The model also allowed for the definition of a figure of merit for magnetostrictive materials that was correlated to the magnetic transduction efficiency. The results further indicated that large magnetic transduction efficiencies close to 100\%{} can be obtained in such magnetoelectric transducers when $Q$-factors on the order of 1000 can be reached. Nonetheless, efficiencies of several 10\%{} can already be obtained even for $Q \sim 10$ to 100 for 20 nm thick Ni or Terfenol-D magnetostrictive layers. Even higher efficiencies can be achieved for thicker films. By comparison, magnetic transduction efficiencies are lower for CoFeB due to a combination of weak magnetostriction and large saturation magnetization, despite low Gilbert damping. 

Together with the favorable scaling behavior, \emph{i.e.} an area-independent efficiency (as long as edge and corner effects can be neglected), the high calculated efficiencies confirm the promise of magnetoelectric resonators as magnetic transducers to excite ferromagnetic resonance or spin waves \cite{Khitun_2007,khitun_magnetoelectric_2009,Khitun11}. The results also clearly demonstrate the challenge of material selection for such devices. When damping does not play a role, both Ni and Terfenol-D appear similarly appealing. For usage as spin-wave waveguide materials however, the Gilbert damping of Ni and Terfenol-D is too large for long distance propagation and other materials are required. The above model can then be used as a guideline to select optimized materials and compare their performance with,\emph{e.g.}, low-$\alpha$ CoFeB studied here.

\appendix
\section{Magnetoelastic stress and body force}

The magnetoelastic energy for a material with cubic (or higher) crystal symmetry is given by \cite{Gurevich96,Tucker1972}
\begin{equation}
\label{eq:E_mel}
E_\mathrm{mel} =\int\limits_V \left[ \frac{B_1}{M_S^2} \sum_i M_\mathrm{i}^2 S_{ii} + B_2 \sum_{i\neq j} M_{i}M_{j} S_{ij}\right] dV\,.
\end{equation}
\noindent Here, $S_{ij}$ are the strain tensor components and $B_{1,2}$ are the magnetoelastic coupling constants. The magnetoelastic stress tensor then becomes \cite{Vanderveken2021}
\begin{equation}
	\sigma_{\mathrm{mel},ij} =  \frac{\delta E_\mathrm{mel}}{\delta S_{ij}} \,.
\end{equation}
Considering only the shear strain $S_{xz}=S_{zx}$ to be nonzero---as it is the case in the configuration in this work---results in nonzero shear stress $\sigma_{zx}=\sigma_{xz}$ only. This stress component is then given by 
\begin{equation}
\sigma_{\mathrm{mel},xz} = \frac{B_2}{M_S^2} M_xM_z\,.
\end{equation}
The body force is found by taking the derivative of the stress tensor, leading to\cite{Vanderveken2021, Graff1975,Achenbach1973}
\begin{equation}
\label{eq:bf}
	f_{\mathrm{mel},i} = \frac{\partial}{\partial x_{j}} \frac{\delta E_\mathrm{tot}}{\delta S_{ij}}\,.
\end{equation}

\section{Magnetization dynamics as a function of strain}

In this appendix, the amplitude of the magnetization dynamics is determined as a function of the strain amplitude. The action of the strain onto the magnetization is captured by the magnetoelastic field, which is given by
\begin{align}
\label{eq:Hmel_gen}
    \mathbf{H}_{\mathrm{mel}} &= -\frac{1}{\mu_0} \frac{\delta E_\mathrm{mel}}{\delta\mathbf{M}} \\
    &= -\frac{2}{\mu_0M_\mathrm{s}^2}  \begin{bmatrix}
B_1S_{{xx}}M_x + B_2(S_{{xy}}M_y+S_{zx}M_z) \\
B_1S_{{yy}}M_y + B_2(S_{{xy}}M_x+S_{yz}M_z) \\
B_1S_{{zz}}M_z + B_2(S_{{zx}}M_x+S_{yz}M_y) 
\end{bmatrix}\, .
\end{align}
\noindent For the considered geometry and configuration, only the shear strain $S_{xz}=S_{zx}$ is present, which results in
\begin{equation}
\label{eq:Hmel}
    \mathbf{H}_{\mathrm{mel}} = -\frac{2B_2S^m_{{zx}}}{\mu_0M_\mathrm{s}^2}  \begin{bmatrix}
 M_z \\
 0 \\
 M_x 
\end{bmatrix}\,.
\end{equation}
Below, the subscripts are removed to improve readability, \emph{i.e.}~$B_2 \equiv B$ and $S^m_{zx} \equiv S^m$. 

The other contributions to the effective magnetic field include the exchange field, the demagnetization field, and the external applied field $H_0$. However, assuming that the magnetic layer is much thinner than the acoustic wavelength, \emph{i.e.} $k_mt\ll 1$, there is only a weak strain gradient along the magnetic layer thickness. Hence, the magnetic excitations can be considered as a uniform along the magnetostrictive film thickness with very weak magnetization variation in this direction. This allows for the neglection of the exchange interaction. The demagnetization field can then be approximated by
\begin{equation}
H_d = \begin{bmatrix}
    0 \\
    0 \\
    -M_z
    \end{bmatrix} \,.
\end{equation} 
Adding the external, demagnetization, and magnetoelastic fields results in the effective field 
\begin{equation}
    H_\mathrm{eff} = \begin{bmatrix}
    H_0 -\frac{2BS^m}{\mu_0M_\mathrm{s}^2}M_z \\
    0 \\
    -M_z-\frac{2BS^m}{\mu_0M_\mathrm{s}^2} M_x
    \end{bmatrix} \,.
\end{equation}
Inserting this effective field into the LLG equation and linearizing this differential equation results in
\begin{equation}
    i\omega \begin{bmatrix} M_y \\ M_z \end{bmatrix} = \begin{bmatrix} -(\omega_0+\omega_M)M_z-2BS^m\gamma \\ \omega_0 M_y \end{bmatrix} + i\omega \alpha \begin{bmatrix} -M_z \\M_y \end{bmatrix}\,.
\end{equation}
These equations can be rewritten to find the relations between the magnetization components and the strain
\begin{align}
M_y &= -i\frac{2B\gamma  \omega}{\omega_y\omega_z-\omega^2} S^m \\
&= -i\frac{2B\gamma  \omega}{\omega_r^2-\omega^2 +i\alpha \omega(2\omega_0+\omega_M)} S^m
\end{align}
\begin{align}
M_z &= -\frac{2B\gamma  \omega_y}{\omega_y\omega_z-\omega^2} S^m \\
&= -\frac{2B\gamma  \omega_y}{\omega_r^2-\omega^2 +i\alpha \omega(2\omega_0+\omega_M)} S^m
\end{align}
with $\omega_y = \omega_0+i\omega\alpha$ and $\omega_z = \omega_0+\omega_M+i\omega\alpha$. Assuming that $\alpha\ll 1$, we can write
\begin{equation}
    \omega_y\omega_z \approx \omega_0(\omega_0+\omega_M)+i\omega\alpha(2\omega_0+\omega_M) = \omega_r^2+i\omega\alpha(2\omega_0+\omega_M)
\end{equation}
\noindent with $\omega_r^2 = \omega_0(\omega_0+\omega_M) $ the magnetic resonance frequency. 
Note that the dynamic magnetization has the same weak thickness variation as the strain $S^m(z)$.

\begin{acknowledgments}

This work has been supported by imec’s industrial affiliate program on beyond-CMOS
logic. It has also received funding from the European Union’s Horizon 2020 research and
innovation program within the FET-OPEN project CHIRON under grant agreement No.
801055. F.V. acknowledges financial support from the Research Foundation -- Flanders
(FWO) through grant No. 1S05719N. The authors would also like to thank Roman Verba for useful discussions. 

\end{acknowledgments}

\bibliography{refs}

\end{document}